\documentclass[12pt]{article}%
\usepackage[nosort]{cite}
\usepackage{graphicx}
\usepackage{multicol}
\usepackage{amsfonts}
\usepackage{amssymb}
\usepackage{amsmath}
\usepackage{heck}
\usepackage[all]{xy}
\usepackage{setspace}
\usepackage{verbatim}
\usepackage{color}
\usepackage{epsfig}
\usepackage[margin=1in]{geometry}
\usepackage{titletoc}
\usepackage{lscape}
\usepackage{tikz}
\usepackage[debug,pageanchor=false]{hyperref}%
\providecommand{\U}[1]{\protect\rule{.1in}{.1in}}
\numberwithin{equation}{section}

\definecolor{link}{rgb}{.8,.15,.1}
\hypersetup{colorlinks=true,linkcolor=link,citecolor=link,urlcolor=link,linktocpage}
\begin{document}

\date{March 2015}
\title{On the Defect Group of a 6D SCFT}

\institution{HARVARD}{\centerline{${}^{1}$ Jefferson Physical Laboratory, Harvard University, Cambridge, MA 02138, USA}}\institution{UNC}{\centerline{${}^{2}$ Department of Physics, University of North Carolina, Chapel Hill, NC 27599, USA}}
\institution{SCGP}{\centerline{${}^{3}$ Simons Center for Geometry and Physics, Stony Brook University, NY 11794, USA}}

\authors{Michele Del Zotto\worksat{\HARVARD}\footnote{e-mail: {\tt delzotto@physics.harvard.edu}},
Jonathan J. Heckman\worksat{\UNC}\footnote{e-mail: {\tt jheckman@email.unc.edu}},
\\[4mm] Daniel S. Park\worksat{\SCGP}\footnote{e-mail: {\tt dpark@scgp.stonybrook.edu}},
and Tom Rudelius\worksat{\HARVARD}\footnote{e-mail: {\tt rudelius@physics.harvard.edu}}}

\abstract{We use the F-theory realization of 6D superconformal field theories (SCFTs) to study the corresponding spectrum
of stringlike, i.e. surface defects. On the tensor branch, all of the stringlike excitations pick up a finite tension, and there is a corresponding lattice of string charges, as well as a dual lattice of charges for the surface defects. The defect group
is data intrinsic to the SCFT and measures the surface defect charges which are not screened by dynamical strings. When non-trivial, it indicates that the associated theory has a partition vector rather than a partition function. We compute the defect group for all known 6D SCFTs, and find that it is just the abelianization of the discrete subgroup of $U(2)$ which appears in the classification of 6D SCFTs realized in F-theory. We also explain how the defect group specifies defining data in the compactification of a $(1,0)$ SCFT.}

\maketitle


\enlargethispage{\baselineskip}

\setcounter{tocdepth}{2}

\newpage

\section{Introduction \label{sec:INTRO}}

An outstanding open problem of theoretical physics is to fully characterize the
defining data of a quantum field theory. One important insight from recent
work on non-perturbative aspects is that specifying
the spectrum of local operators is typically not enough to complete this
characterization. Rather, additional information involving the spectrum of extended
objects must also be included. For example, in the context of four-dimensional
theories, the spectrum of line operators and surface operators must be specified \cite{VafaWitten,Ganor:1996nf,Kapustin:2005py,Gaiotto:2010be,Aharony:2013hda, Gaiotto:2014kfa}.
A notable example of this type are the 4D $\mathcal{N}=2$ theories of class $\mathcal{S}$
\cite{Witten:1997sc,Gaiotto:2009we,Alday:2009aq,Gaiotto:2009hg} obtained
by compactifications of the
6D\ $(2,0)$ theories on Riemann surfaces
\cite{Witten:2009at,Drukker:2009tz,Alday:2009fs,Tachikawa:2013hya,Xie:2013vfa}.

In this note we begin the study of extended defects for six-dimensional
superconformal field theories (SCFTs) with minimal supersymmetry.\footnote{The role of the global structure
of extended defects in six-dimensional supergravity theories
was discussed in \cite{Seiberg:2011dr}.}
Recently, steady progress on the classification
of such theories has been made \cite{Heckman:2013pva, Gaiotto:2014lca, DelZotto:2014hpa,
Heckman:2014qba, DelZotto:2014fia, Heckman:2015bfa, Bhardwaj:2015xxa}, with a recently announced
classification of theories (which is quite possibly complete) which can arise from compactifications
of F-theory \cite{Heckman:2015bfa}. A hallmark of these systems is the existence of
tensionless strings in the low energy effective field theory. As these are
extended objects, it is natural to also expect a physically rich spectrum of
defects. For earlier work on realizing 6D SCFTs in string theory, see e.g. \cite{Witten:1995ex,
Witten:1995zh, Strominger:1995ac, WittenSmall,
Ganor:1996mu,MorrisonVafaII,Seiberg:1996vs, Seiberg:1996qx, Bershadsky:1996nu,
Brunner:1997gf, Blum:1997fw, Aspinwall:1997ye, Intriligator:1997dh,
Hanany:1997gh}.

In the F-theory realization of these theories, we have an elliptically fibered
Calabi-Yau threefold $X\rightarrow B$ over a non-compact base. The tensionless
strings arise from D3-branes wrapping contracting spheres of $B$, and the
associated lattice of string charges is $\Lambda_{\text{string}}%
=H_{2}^{\text{cpct}}(B,%
\mathbb{Z}
)$. Geometrically, the dual lattice $\Lambda_{\text{string}}^{\ast}$ is
associated with a basis of non-compact two-cycles canonically
paired with elements of $\Lambda_{\text{string}}$. Wrapping a D3-brane over
such a non-compact cycle leads to a surface defect in the effective 6D
theory. Physically then, we can view $\Lambda_{\text{string}}$ as the lattice of charges for
stringlike excitations and $\Lambda_{\text{string}}^{\ast}$ as the lattice of charges
for stringlike defects, i.e. ``surface operators''. This singles out the unimodular SCFTs, i.e. those
theories with $\Lambda_{\text{string}} = \Lambda_{\text{string}}^{\ast}$. Unimodularity of the charge lattice is a necessary condition for the given 6D model to have a partition function on curved manifolds (see e.g. \cite{Seiberg:2011dr, DelZotto:2014fia, Gaiotto:2014kfa}).

But in general, there can be a mismatch between the lattices $\Lambda
_{\text{string}}$ and $\Lambda_{\text{string}}^{\ast}$, and this mismatch is
measured by the abelian \textquotedblleft defect group\textquotedblright\ of
the theory:%
\begin{equation}
\mathcal{C} \equiv \Lambda_{\text{string}}^{\ast}/\Lambda
_{\text{string}}.
\end{equation}
This group also determines the obstruction to defining a partition function
for the 6D theory\cite{Seiberg:2011dr}---when $\mathcal{C}$ is non-trivial, the 6D theory is best thought of as
a relative quantum field theory \cite{Witten:2009at,Tachikawa:2013hya,Witten:1996hc,
Aharony:1998qu,Witten:1998wy,Moore:2004jv,Belov:2006jd,Freed:2006yc,Witten:2007ct,
Henningson:2010rc,Freed:2012bs}.

The physical interpretation of $\mathcal{C}$ is the natural generalization of the discrete electric / magnetic 't Hooft fluxes present in non-abelian gauge theory \cite{HooftI, HooftII}, but now in the context of a 6D theory. Recall that for gauge theory, we can move to a phase where the effects of the non-abelian force carriers (i.e. the gluons) are screened at long distances. Then, the spectrum of defects, i.e. the line operators organize according to ``$N$-ality'', i.e. the discrete charge carried by the center of the gauge group. Though it is still an open problem to determine a formulation of the tensionless non-abelian force carriers for a 6D SCFT, we do know that on the tensor branch their effects are screened. What the defect group measures is the charges of surface operators which cannot be screened by the dynamical strings of our theory.

This discrete data shows up most clearly when we consider our 6D theory on a spacetime other than $\mathbb{R}^{5,1}$. In flat space, the locations of defects correspond to deleting certain 2D subspaces from the spacetime. More generally, we can work on a six-manifold $M_6$. The choice of a background flux sector is captured by the cohomology $H^{3}(M_6,\mathcal{C})$.
Then, for any given model each choice of a maximal set of mutually local observables
$L \subset H^{3}(M_6,\mathcal{C})$ corresponds to a set of generalized conformal blocks $Z_{a}$ indexed by
$a\in H^{3}(M_6,\mathcal{C})/L$. For the analogous statement in the case of the 6D (2,0) theories, see e.g. references \cite{Witten:2009at,Tachikawa:2013hya,Witten:1998wy,Henningson:2010rc,Henningson:2009gv}.

In this note we compute the defect group for all known 6D SCFTs. An important result from this computation is that it does not involve a specific choice of resolution of the associated F-theory geometry. Rather, we find that for F-theory on a base obtained by blowing up $\mathbb{C}^2 / \Gamma$ for $\Gamma$ an appropriate discrete subgroup of $U(2)$, the defect group is always the abelianization of $\Gamma$:
\begin{equation}
\mathcal{C} = \mathrm{Ab}[\Gamma].
\end{equation}
All 6D SCFTs which can be generated in F-theory arise from blow ups of $\mathbb{C}^2 / \Gamma$, and possibly making the elliptic fiber more singular. Further, the explicit list of possible $\Gamma \subset U(2)$ which can appear has been worked out in the generalized ADE classification of reference \cite{Heckman:2013pva}. The fact that the defect group is independent of the geometric resolution parameters indicates that this data is intrinsic to the SCFT itself (i.e. it does not depend on the specific values of the string tensions). Summarizing our findings, the class of defect groups we uncover are as follows:%
\begin{equation}%
\begin{tabular}
[c]{|c|c|c|c|c|c|c|}\hline
& A-type & D-type$_{\text{(even)}}$ & D-type$_{\text{(odd)}}$ & $E_{6}$ &
$E_{7}$ & $E_{8}$\\\hline
$\mathcal{C}$ & $\mathbb{Z}_{p}$ & $%
\mathbb{Z}
_{2}\times%
\mathbb{Z}
_{2p}$ & $%
\mathbb{Z}
_{4p}$ & $%
\mathbb{Z}
_{3}$ & $%
\mathbb{Z}
_{2}$ & $1$\\\hline
\end{tabular}
\ \ \ ,
\end{equation}
where there turn out to be two D-type defect groups. Let us note that in the
special case of the $(2,0)$ theories, $\mathcal{C}$ is nothing
but the center of the corresponding simply connected Lie
group.\footnote{Indeed, the center of the Lie group is given by the quotient
of the weight lattice by the root lattice, i.e. $\Lambda_{\text{weight}%
}/\Lambda_{\text{root}}$, and its first homotopy group is given by the
quotient of the co-weight lattice by the co-root lattice, i.e. $\Lambda
_{\text{weight}}^{\vee}/\Lambda_{\text{root}}^{\vee}$. See e.g. references
\cite{BUMP, YELLOW}.}

The rest of this note is organized as follows. In
section \ref{sec:DEFECT} we discuss in more detail
the defect group of a 6D SCFT. In section \ref{sec:BLOWLATT}, we
review some elements of the classification of 6D\ SCFTs, and in
particular the correspondence with generalized ADE\ subgroups of $U(2)$.
In this same section we show that for a given F-theory model, performing
blowups / blowdowns of a base does not change the defect group. We then
proceed in section \ref{sec:DEFECTO} to an explicit determination of all
possible defect groups, and show that it always corresponds to the
abelianization of an associated discrete subgroup of $U(2)$. In section
\ref{sec:COMPACT} we discuss the remnants of this discrete three-form flux
data upon compactification to a lower-dimensional theory. We present our
conclusions in section \ref{sec:CONC}.

\section{The Defect Group \label{sec:DEFECT}}

In this section we introduce the notion of a defect group, and explain in more detail why we expect it to capture
important discrete data of a 6D SCFT. As we shall be relying on their geometric characterization,
let us first recall some additional details on the F-theory construction of a 6D SCFT.

Recall that in F-theory, we
work with a non-compact elliptically fibered Calabi-Yau threefold
$X\rightarrow B$, in which the base $B$ is a non-compact K\"{a}hler surface.
Recent work in \cite{Heckman:2015bfa} has led to a classification of all
possible base geometries $B$, as well as all possible elliptic fibrations over
a given base.

The charge lattice for dynamical strings is captured by $\Lambda
_{\text{string}}=H_{2}^{\text{cpct}}(B,%
\mathbb{Z}
)$, and comes equipped with an intersection pairing, which in turn defines a
natural dot product. Indeed, given a basis of two-cycles $e_{i}=[\Sigma_{i}]$
for $i=1,...,r$, we can then speak of the adjacency matrix for the lattice:%
\begin{equation}
A_{ij}=e_{i}\cdot e_{j}=-\Sigma_{i}\cap\Sigma_{j}\text{,}%
\end{equation}
in the obvious notation. We can also introduce the dual lattice $\Lambda
_{\text{string}}^{\ast}$ with generators $w^{i}$ such that $w^{i}\cdot
e_{j}=\delta_{\text{ \ }j}^{i}$. Geometrically, these generators are
associated with the non-compact two-cycles dual to the compact ones. Physically,
this symmetric pairing is just the Dirac pairing in six dimensions \cite{Deser:1997se}. The fact that $\Lambda^{\ast}_{\mathrm{string}}$ is the lattice of defect charges follows by requiring these are labeled by a maximal set consistent with Dirac quantization. Surface defects are thus labeled by their charge, the surface they wrap and the supersymmetry they preserve.

Let us now turn to the F-theory realization of these defects. Consider a D3-brane wrapped over such a non-compact two-cycle. In the six-dimensional effective theory, we get a non-dynamical effective string with formally infinite tension (as it wraps a non-compact cycle). We refer to this as a surface operator. This should be viewed as a heavy probe of the 6D theory, and is the higher-dimensional generalization of a line operator. Now, in the context of 4D gauge theory, it is well-known that the lattice of defect charges can be screened, leaving us with just the center of the gauge group \cite{HooftI, HooftII}. Though we do not have the 6D analogue of non-abelian gauge theory for two-form potentials, we can still determine the effects of screening using the geometric characterization of these theories.\footnote{We thank C. Vafa for helpful comments.}

Along these lines, let us first briefly discuss the effects of screening for 4D gauge theory from the perspective of geometric engineering. For concreteness, consider type IIB string theory on $T^2 \times \mathbb{C}^{2} / \Gamma_{ADE}$ for $\Gamma_{ADE}$ a discrete ADE subgroup of $SU(2)$. In the four uncompactified directions, we get 4D $\mathcal{N} = 4$ Super Yang-Mills theory with gauge group $G_{ADE}$ the simply connected Lie group of ADE type. Now, one way to understand the non-abelian gauge bosons is to take the small resolution of this singularity. Wrapping D3-branes over the compact two-cycles associated with the roots of the corresponding algebra and the A-cycle of the $T^2$, we see explicitly all the massive states which are going to assemble into the gluons of the theory. Introducing the root lattice $\Lambda_{\mathrm{root}} = H_{2}^{cpct}(B, \mathbb{Z})$, the dual lattice $\Lambda_{\mathrm{root}}^{\ast}$ tells us about the possible line operators in this theory. Precisely because the force carriers are massive, there is a screening of the charge of these line operators. The charges which cannot be screened are simply $\Lambda_{\mathrm{root}}^{\ast} / \Lambda_{\mathrm{root}}$. In group theoretic terms, it is also known that this quotient is nothing but $\mathcal{Z}(G_{ADE})$, the center of the corresponding simply connected Lie group of ADE type. Observe that in the special case $G = SU(N)$, the center is the $N$-ality group $\mathbb{Z}_N$.

Consider next the case of our 6D SCFTs. Here, we do not have a formulation of interacting non-abelian two-form potentials. Nevertheless, from the geometry, we can still see that on the tensor branch, the associated string-like excitations have picked up a non-zero tension. Indeed, the reason it is possible to say anything concrete at all about the tensor branch is that the associated ``off-diagonal'' force carriers are necessarily screened (as they have picked up a tension). From this perspective, we can follow the same analysis as we did for the 4D gauge theory: We take the lattice of compact two-cycles $\Lambda_{\mathrm{string}}$, introduce its dual $\Lambda_{\mathrm{string}}^{\ast}$ for the surface defects, and the quotient group:
\begin{equation}
\mathcal{C} = \Lambda^{\ast}_{\mathrm{string}} / \Lambda_{\mathrm{string}}
\end{equation}
tells us what charges for the surface operators cannot be screened.

Inserting a surface defect in six dimensions can be viewed as deleting its 2D worldvolume from our 6D spacetime. This is why the global topology
of the spacetime figures into our discussion of surface defects. On a six-manifold $M_6$, the discrete three-form fluxes are given by $H^{3}(M_6 , \mathcal{C}) \simeq H^{3}(M_6 , \mathbb{Z}) \otimes \mathcal{C}$.\footnote{Throughout, we assume that $H^3 (M_6,\mathbb{Z})$ does not have a torsion group. Issues related to the torsion of $H^3 (M_6,\mathbb{Z})$ have been discussed at length, for example, in \cite{Witten:1998wy}.}

When $\mathcal{C}$ is non-trivial, the 6D
theory is a relative quantum field theory, rather than a conventional
quantum field theory. Much of what we say in the rest of this section has
been stated in various forms in the literature for the $(2,0)$
SCFTs and, for that matter, relative quantum field theories in general
 \cite{Witten:2009at,Tachikawa:2013hya,
Witten:1996hc,Aharony:1998qu,Witten:1998wy,
Moore:2004jv,Belov:2006jd,Freed:2006yc,Witten:2007ct,
Henningson:2010rc,Freed:2012bs}.
The main novelty
here is to show how these same structures naturally persist in the context of
the $(1,0)$ theories.

The theory on the background Euclidean spacetime $M_6$ is not expected to have a well-defined partition function.
Rather, it has a collection of ``partition functions" that can be organized into
a ``partition vector." It is natural to speculate that elements of this vector space can be understood as
states in the Hilbert space of some 7D topological
field theory, as has been worked out for some of the $(2,0)$ theories.
The Hilbert space is then given by an irreducible representation
of a particular Heisenberg group, which we now define.

The Heisenberg group is defined by noticing that there is a
natural pairing $E:H^{3}(M_{6},\mathcal{C})\times H^{3}(M_{6},\mathcal{C}%
)\rightarrow U(1)$ which takes $h,h^{\prime}\in H^{3}(M_{6},\mathcal{C})$ to
$\exp(2 \pi i\left\langle h,h^{\prime}\right\rangle )$, in the obvious notation \cite{Tachikawa:2013hya}.%
\footnote{This pairing is derived from a canonical pairing
$\mathcal{C} \times \mathcal{C} \rightarrow U(1)$ for elements of
group $\mathcal{C}$. For example, when $\mathcal{C}=\mathbb{Z}_N$,
the pairing is given by
\begin{equation}
\exp(2 \pi i \langle n , m \rangle) =
\exp \left({2 \pi i nm \over N} \right) \,.
\end{equation}
Notice that it is convenient to present $\mathcal{C}$
as an additive abelian group for such purposes---we shall use this same convention in section \ref{sec:COMPACT}.}
This enables us to define an extension, the Heisenberg group
$\underline{H}^{3}(M_{6},\mathcal{C})$ via the exact sequence:%
\begin{equation}
1\rightarrow U(1)\rightarrow\underline{H}^{3}(M_{6},\mathcal{C})\rightarrow
H^{3}(M_{6},\mathcal{C})\rightarrow0.
\end{equation}
In more practical terms, to any
given element $h\in H^{3}(M_{6},\mathcal{C})$, we can assign a corresponding quantum flux  $\Phi
(h)\in\underline{H}^{3}(M_{6},\mathcal{C})$ \cite{Freed:2006yc}. These elements are subject to the
comutation relation:%
\begin{equation}
\Phi(h)\Phi(h^{\prime})=\exp(2 \pi i\left\langle h,h^{\prime}\right\rangle
)\Phi(h^{\prime})\Phi(h).
\end{equation}

By a famous theorem by Stone, von Neumann and Mackey,
the Heisenberg group $\underline{H}^{3}(M_{6},\mathcal{C})$ has a unique
irreducible representation (see, e.g., \cite{HeisenbergGroup,TataTheta}).
The vector space of this representation is
the Hilbert space of the seven-dimensional topological quantum field theory,
or equivalently, the ``partition vector space." This vector space can be
built out of a maximal isotropic subgroup $L$ of
${H}^{3}(M_{6},\mathcal{C})$ \cite{Tachikawa:2013hya,Witten:1998wy,
HeisenbergGroup}. First, there is a unique ray in the vector space,
which we represent by the vector $Z^L_0$ that is invariant under the
action of $L$. Denoting the coset
\begin{equation}
L^\perp \equiv {H}^{3}(M_{6},\mathcal{C})/L \,,
\end{equation}
the rest of the basis vectors are obtained by acting on $Z^L_0$ by
elements $v \in L^\perp$:
\begin{equation}
Z^L_v = \Phi(v) Z^L_0 \,.
\end{equation}
Note that $Z^L_v$ are eigenvectors for the elements $\Phi(w)$ with
$w \in L$:
\begin{equation}
\Phi(w) Z^L_v = \exp(2 \pi i \langle w,v \rangle) Z^L_v  \,.
\label{basis}
\end{equation}
In particular, while the overall normalization of $Z^L_0$ is not fixed,
the normalization of $Z^L_v$ is fixed with respect to $Z^L_0$.
The theorem of Stone, von Neumann and Mackey states that given
two maximal isotropic subgroups $L$ and $L'$, the two representations
constructed this way are isomorphic. In particular, there is an invertible
linear transformation $R$ such that for all $v' \in {L'}^\perp$,
\begin{equation}
Z^{L'}_{v'} = \sum_{v \in L^\perp} R_{v'}^{~v} Z^L_{v} \,.
\end{equation}
In physical terms,
we must demand that our partition vector can be decomposed into eigenfunctions
of discrete fluxes which can all be simultaneously measured, the condition of mutal locality of the fluxes
being encoded in the pairing $E:H^{3}(M_{6},\mathcal{C})\times H^{3}(M_{6},\mathcal{C}%
)\rightarrow U(1)$. That means we must pick a sublattice
$L\subset H^{3}(M_{6},\mathcal{C})$ such that for all pairs of elements $h,h^\prime \in L$, $E(h,h^\prime) = 1$. This is precisely the definition of an isotropic subgroup of $H^{3}(M_{6},\mathcal{C})$. The physical
interpretation of $L$ is that it defines a maximal collection of mutually local
discrete fluxes.\footnote{It is helpful to recall the distinction between
gauge theory with group $SU(N)$ and $SU(N)/%
\mathbb{Z}
_{N}$. In the latter case, we impose a restriction on the admissible Wilson
lines we consider, i.e. we restrict the possible representations, and in
particular exclude the fundamental representation.} We stress that any maximally
mutually local sublattice $L$ would work, and each such choice gives rise to a
different realization of the generalized conformal blocks of the theory.

\section{Defects and Generalized ADE \label{sec:BLOWLATT}}

Having introduced the defect group of a 6D SCFT, we now determine some of its general properties.
The main result from this section will be that the generalized ADE
classification of 6D SCFTs (with an F-theory realization) in terms of specific
discrete subgroups $\Gamma \subset U(2)$ is enough to determine the list of
possible defect groups. In geometric terms, this is the statement that in
F-theory, all of the bases are obtained from blowups of $\mathbb{C}^2 / \Gamma$ and
that the defect group is invariant under such blowups. A corollary of this
result is that the data of the defect group is insensitive to the geometric resolution parameters and thus to the
specific tensions for our effective strings, as one should expect for data intrinsic to the SCFT itself.

To frame the discussion to follow, we review some elements of how to build and
classify 6D\ SCFTs in F-theory, and in particular the correspondence with
generalized ADE\ discrete subgroups of $U(2)$. One of the results from reference \cite{Heckman:2015bfa} is a full
classification of possible bases. These are composed of configurations of
$\mathbb{P}^{1}$'s with self-intersection $-n$ with $1\leq n\leq12$.
Importantly, each $-1$ curve intersects at most two other curves. On the
diagonal of the adjacency matrix $A_{ij}$, the entries are all positive, and
the only off-diagonal entries are $0$ or $-1$ and the structure of the graph
does not contain any closed loops. Further, it turns out that the resulting
structures of possible configurations of curves are remarkably constrained,
and have the form of a long \textquotedblleft spine\textquotedblright\ with a
small amount of decoration on the ends \cite{Heckman:2015bfa}.

The building blocks for these bases involve the so-called non-Higgsable
clusters (NHCs) of reference \cite{Morrison:2012np}, which include the
important observation that there is a minimal canonical singular fiber type
associated with each such cluster. The classification result of
\cite{Heckman:2015bfa} also involves determining possible enhancements in the
elliptic fibration over a given base. Observe that if the base is fixed, this
additional data does not affect the lattice of string charges. It will
therefore not figure in to our considerations.

Given a base, a natural geometric operation is to consider the blowdown of all
$-1$ curves in the geometry. This shifts the self-intersection numbers of
neighboring curves according to the rule:%
\begin{equation}
(n+1,1,m+1)\rightarrow(n,m).
\end{equation}
Continuing in this way, we can iteratively blow down all $-1$ curves until we
reach an ``endpoint'' geometry in which all curves have self-intersection $-2$ or less.
Quite remarkably, it turns out that these geometries are minimal resolutions of the orbifold singularities
$\mathbb{C}^{2}/\Gamma$ where $\Gamma$ is a discrete subgroup of $U(2)$.\footnote{Note
that to define an F-theory background, further blowups in the base are required.} One of the results of
reference \cite{Heckman:2013pva} is that there is a generalized ADE\ classification of $(1,0)$
theories. Their endpoint configurations are:
\begin{align}
\text{A-type Endpoint} &  \text{: \ \ }n_{1},...,n_{k}\label{ATYPE}\\
\text{D-type Endpoint} &  \text{: \ \ }2,\overset{2}{m_{1}},...,m_{l}\\
\text{E-type Endpoint} &  \text{: \ \ }\left\{
\begin{array}
[c]{l}%
2,2,\overset{2}{2},2,2\\
2,2,\overset{2}{2},2,2,2\\
2,2,\overset{2}{2},2,2,2,2
\end{array}
\right\}  ,\label{ETYPE}%
\end{align}
where the $n_{i}$ and $m_{j}$ correspond to curves of self-intersection
$-n_{i}$ and $-m_{j}$, in the obvious notation. Not all values of the $n$'s and $m$'s occur,
but the full list which do appear can be found in reference \cite{Heckman:2013pva}. Note that in the context of
F-theory, the E-type subgroups are always a discrete subgroup of $SU(2)$.
Finally, we can of course also have bases which blow down to $\mathbb{C}^{2}$.
For example a configuration of curves such as $1,2,...,2$ is of this type.

Now, given an endpoint there is always a minimal set of blowups required to
reach a consistent F-theory model. Additionally, other \textquotedblleft
unforced blowups\textquotedblright\ can sometimes be included. In more detail,
the operation of blowing up the intersection point involves the following
procedure. Given curves $\Sigma_{L}$ and $\Sigma_{R}$ of self-intersection
$-n$ and $-m$ which intersect at one point, blowing up this point introduces a
new exceptional divisor $\Sigma_{\text{new}}$ and shifts the homology classes
as follows:%
\begin{equation}
\lbrack\Sigma_{L}]\rightarrow\lbrack\Sigma_{L}]+[\Sigma_{\text{new}}]\text{
\ \ and \ \ }[\Sigma_{R}]\rightarrow\lbrack\Sigma_{R}]+[\Sigma_{\text{new}}].
\end{equation}

Let us now show that blowing up the base does not alter the group $\Lambda_{\text{string}%
}^{\ast}/\Lambda_{\text{string}}$. To establish this, we proceed inductively.
Suppose that we have a lattice $\Lambda_{\text{string}}$, and that we then
perform another blowup on a point of one of the curves of our base. This
introduces an additional exceptional divisor $\Sigma_{\text{new}}$, so the new
lattice of string charges $\Lambda_{\text{string}}^{(1)}$ has increased in
rank by one. Additionally, a divisor class which touches the new curve will
shift as $[\Sigma]\rightarrow\lbrack\Sigma]+[\Sigma_{\text{new}}]$.

In fact, by an integral change of basis, we see that the new lattice of string
charges is really just $\Lambda_{\text{string}}^{(1)}\simeq\Lambda
_{\text{string}}\oplus%
\mathbb{Z}
$, where the additional factor is generated by the new class $[\Sigma_{\text{new}}]$.
Since $[\Sigma_\text{new}]$ has self-intersection $-1$,
this additional factor is a one-dimensional self-dual lattice. By a
similar token, if we consider $b$ blowups of a base, we reach a lattice
$\Lambda_{\text{string}}^{(b)}$ which is:%
\begin{equation}
\Lambda_{\text{string}}^{(b)}\simeq\Lambda_{\text{string}}\oplus
\underset{b}{\underbrace{%
\mathbb{Z}
\oplus...\oplus%
\mathbb{Z}
}}.
\end{equation}
Now, because the lattice $%
\mathbb{Z}
^{\oplus b}$ is already self-dual, the dual lattice is:%
\begin{equation}
\Lambda_{\text{string}}^{(b)\ast}\simeq\Lambda_{\text{string}}^{\ast}%
\oplus\underset{b}{\underbrace{%
\mathbb{Z}
\oplus...\oplus%
\mathbb{Z}
}}.
\end{equation}
So, we have the important fact that the defect group is unchanged by the
number of blowups:%
\begin{equation}
\Lambda_{\text{string}}^{(b)\ast}/\Lambda_{\text{string}}^{(b)}\simeq
\Lambda_{\text{string}}^{\ast}/\Lambda_{\text{string}}\text{ \ \ for all
}b\text{.}%
\end{equation}

Recall, however, that all of the F-theory bases for 6D\ SCFTs arise from
blowups of generalized ADE orbifolds of $\Gamma\subset U(2)$. What this means
is that for the purposes of calculating $\Lambda_{\text{string}}^{\ast
}/\Lambda_{\text{string}}$, we can confine our analysis to the minimal
resolution of the orbifold singularity $\mathbb{C}^{2}/\Gamma$, as described
by lines (\ref{ATYPE})-(\ref{ETYPE}). To sum up, we see that for the
purposes of determining the defect group $\Lambda_{\text{string}}^{\ast
}/\Lambda_{\text{string}}$, it is enough to focus on the geometry of
$\mathbb{C}^{2}/\Gamma$.

\section{Defect Groups of $(1,0)$ Theories \label{sec:DEFECTO}}

In the previous section we established that the generalized ADE-type of a
$(1,0)$ theory determines the defect group $\mathcal{C}%
=\Lambda_{\text{string}}^{\ast}/\Lambda_{\text{string}}$. In this section we
compute the explicit form of $\mathcal{C}$.

To begin, we note that the order of the defect group is readily computed from
the adjacency matrix. The adjacency matrix $A$ defines an embedding of the
lattice $\Lambda_{\text{string}}$ in $\Lambda_{\text{string}}^{\ast}$:%
\begin{equation}
A:\Lambda_{\text{string}}\rightarrow\Lambda_{\text{string}}^{\ast}.
\end{equation}
As a consequence, the order of the defect group satisfies $\left\vert
\mathcal{C}\right\vert =$ $\det A$.

We can also determine the abstract form of this group. For $\Lambda
_{\text{string}}$ a rank $k$ lattice, the group $\Lambda_{\text{string}}%
^{\ast}/\Lambda_{\text{string}}$ is given by the abelian group on $k$
commuting generators $a_{1},...,a_{k}$ subject to the relations:%
\begin{equation}
\overset{k}{\prod_{j=1}}a_{j}^{A_{ij}}=1\label{relator}%
\end{equation}
for $i=1,...,k$. Similar formulae appear in the computation of D-brane spectra
in Landau-Ginzburg vacua. This is of course not an accident, since the
computation of the spectrum of defects is quite parallel to that situation.
Indeed, we will also establish that the defect group is the
abelianization Ab$[\Gamma]$ for the associated discrete group $\Gamma\subset
U(2)$. This is again suggestive of a larger structure in the context of F-theory
backgrounds, on which we comment further in section \ref{sec:CONC}.

Our plan in this section will be to compute the precise form of the defect
groups for all 6D\ SCFTs. To this end, we first compute this data in the case
of an endpoint which is an ADE\ subgroup of $SU(2)$. Here, we observe that
$\mathcal{C}$ is nothing but the center of the corresponding simply connected
ADE\ Lie group. We then establish similar formulae for the generalized A-type
and D-type theories.

\subsection{The ADE Bases}

Our plan in this section will be to consider F-theory models with base an
ADE\ singularity. The minimal resolution of the base gives a bouquet of $-2$
curves which organize according to the respective Dynkin diagram. When the
elliptic fibration is trivial, i.e. when the F-theory background is of the
form $\mathbb{C}^{2}/\Gamma\times T^{2}$ for $\Gamma$ an ADE\ subgroup of
$SU(2)$, we reach a 6D\ $(2,0)$ theory. When the elliptic fibration is
non-trivial over the compact curves, we instead reach a $(1,0)$ theory.

Now, in the special context of the $(2,0)$ theories, it is well known that the
center of the associated simply connected Lie group also determines the
obstruction to specifying a partition function, and thus the defect group. Not
coincidentally, the centers of the associated simply connected Lie group of
the same ADE\ type exactly match the defect group, as well as the
abelianization of the associated discrete subgroup of $SU(2)$:%
\begin{equation}%
\begin{tabular}
[c]{|c|c|c|c|c|c|c|}\hline
& $A_{k}$ & $D_{2k}$ & $D_{2k+1}$ & $E_{6}$ & $E_{7}$ & $E_{8}$\\\hline
$\mathcal{Z}(G_{ADE})=\mathcal{C}(\Gamma_{ADE})=$ \ Ab$[\Gamma_{ADE}]$ &
$\mathbb{Z}_{k+1}$ & $%
\mathbb{Z}
_{2}\times%
\mathbb{Z}
_{2}$ & $%
\mathbb{Z}
_{4}$ & $%
\mathbb{Z}
_{3}$ & $%
\mathbb{Z}
_{2}$ & $1$\\\hline
\end{tabular}
\ \ \ .
\end{equation}
Our plan in this subsection will be to perform a direct computation to verify
the equivalence of the defect group and the abelianization of the quotient group,
which holds for a general F-theory base on an
ADE\ singularity. The material of this subsection collects well-known results,
which are encapsulated by the classical McKay correspondence \cite{MR604577}.
Our main purpose here is to establish notation and set the stage for the computation for all
$(1,0)$ theories.

\subsubsection*{A-type Theories}

Consider first the A-type bases. These are realized in F-theory on a base
consisting of $k$ curves of self-intersection $-2$ arranged as in the
corresponding Dynkin diagram. Returning to our general formula in equation
(\ref{relator}), we have:%
\begin{align}
(a_{1})^{2}  &  =a_{2}\\
(a_{i})^{2}  &  =a_{i-1}a_{i+1}\text{ \ \ for \ \ }1<i<k\\
(a_{k})^{2}  &  =a_{k-1}.
\end{align}
Iteratively solving these constraints, we learn:%
\begin{equation}
a_{i}=(a_{1})^{i}\text{ \ \ and \ \ }(a_{1})^{k+1}=1.
\end{equation}
So in other words, the group is generated by $a_{1}$, an element of order
$k+1$. We therefore learn that $\mathcal{C}(A_{k})=\mathbb{Z}_{k+1}$. Note
that the abelianization Ab$[\Gamma_{A_{k}}]=\mathbb{Z}_{k+1}$, since the group
is abelian.

\subsubsection*{D-type Theories}

Consider next the D-type bases. These are realized in F-theory on a base
consisting of $k$ curves of self-intersection $-2$ arranged as in the
corresponding Dynkin diagram. In this case, it is convenient to observe that
$\det A=4$ for all of the D-type Cartan matrices, so the defect group is
necessarily of order four. Performing a similar computation to that given for
the A-type theories, we learn that $\mathcal{C}$ is either the cyclic group of
order four, or the Klein four-group. The particular case which is realized
depends on whether we have an even number of $-2$ curves or an odd number. For
$k$ even, we get $\mathbb{Z}_{2}\times\mathbb{Z}_{2}$, while for $k$ odd, we
get $\mathbb{Z}_{4}$.

We can also determine that the abelianization Ab$[\Gamma_{D_{k}}]$ matches to
these choices. Recall that the D-type subgroups of $SU(2)$ are the binary
dihedral groups. The binary dihedral group and its abelianization are given by
the following abstract group with two generators:%
\begin{align}
\Gamma_{D_{k}} &  =\left\langle x,a|a^{2k-4}=1\text{, \ \ }x^{2}%
=a^{k-2}\text{, \ \ }xax^{-1}=a^{-1}\right\rangle \\
\text{Ab}[\Gamma_{D_{k}}] &  =\left\langle x,a|a^{2k-4}=1\text{, \ \ }%
x^{2}=a^{k-2}\text{, \ \ }a^{2}=1\right\rangle _{\text{comm}}.
\end{align}
Here, the subscript \textquotedblleft comm\textquotedblright\ amounts to
imposing the relations $gh=hg$ for all elements $g$ and $h$ of the group.
Depending on whether $k$ is even or odd, we get two different abelianizations:%
\begin{equation}
\text{Ab}[\Gamma_{D_{k}}]=\left\{
\begin{array}
[c]{l}%
\mathbb{Z}_{2}\times\mathbb{Z}_{2}\text{ \ \ \ \ \ \ }k\text{ even}\\
\mathbb{Z}_{4}\text{ \ \ \ \ \ \ \ \ \ \ \ \ \ }k\text{ odd}%
\end{array}
\right\}  .
\end{equation}

\subsubsection*{E-type Theories}

Finally, consider the E-type theories. These are realized by F-theory on a
base consisting of $-2$ curves arranged according to the corresponding $E_{6}%
$, $E_{7}$ and $E_{8}$ Dynkin diagrams. To compute the defect group in these
cases, we observe that the determinant of the adjacency matrix is respectively
$3$, $2$, and $1$. We thus learn that the defect groups are:%
\begin{equation}%
\begin{tabular}
[c]{|c|c|c|c|}\hline
& $E_{6}$ & $E_{7}$ & $E_{8}$\\\hline
$\mathcal{C}$ & $%
\mathbb{Z}
_{3}$ & $%
\mathbb{Z}
_{2}$ & $1$\\\hline
\end{tabular}
\ \ \ \ .
\end{equation}
Turning next to the abelianization, we recall that the binary tetrahedral
group (i.e. $E_{6}$), the binary icosahedral group (i.e. $E_{7}$) and the
binary octahedral group (i.e. $E_{8}$) and their respective abelianizations are:%
\begin{align}
\Gamma_{E_{6}} &  =\left\langle a,b,c|a^{3}=b^{3}=c^{2}=abc\right\rangle
\text{ \ \ and \ \ Ab}[\Gamma_{E_{6}}]\mathbb{=}\left\langle \zeta|\zeta
^{3}=1\right\rangle \\
\Gamma_{E_{7}} &  =\left\langle a,b,c|a^{4}=b^{3}=c^{2}=abc\right\rangle
\text{ \ \ and \ \ Ab}[\Gamma_{E_{7}}]\mathbb{=}\left\langle \zeta|\zeta
^{2}=1\right\rangle \\
\Gamma_{E_{8}} &  =\left\langle a,b,c|a^{5}=b^{3}=c^{2}=abc\right\rangle
\text{ \ \ and \ \ Ab}[\Gamma_{E_{8}}]\mathbb{=}\left\langle \zeta
|\zeta=1\right\rangle ,
\end{align}
and as expected, there is an exact match between the two characterizations.

\subsection{Generalized A-type Bases}

Let us \ now turn to the computation of the defect group for the generalized
A-type bases. Recall that these are given by blowups of a configuration of
curves of self-intersection $-n_{1},...,-n_{k}$ which intersect pairwise,
forming a single chain of curves. Contracting these curves leads to an
orbifold singularity $\mathbb{C}^{2}/\Gamma$ where the group action is \cite{jung,MR0062842,Brieskorn,Riemen:dvq}:%
\begin{equation}
(u,v)\rightarrow(\omega u,\omega^{q}v)\text{ \ \ with \ \ }\omega^{p}=1,
\end{equation}
and the integers $p$ and $q$ are determined by the Hirzebruch-Jung continued
fraction:%
\begin{equation}
\frac{p}{q}=n_{1}-\frac{1}{n_{2}-\frac{1}{n_{3}-...\frac{1}{n_{k}}}}.
\end{equation}
Clearing denominators in the continued fraction, we can extract the
corresponding values of $p$ and $q$. These are given by the determinants of
the adjacency matrix $A$, as well as $A^{(1)}$, the matrix obtained by
deleting the first row and column. (i.e. those containing the entry $n_1$):%
\begin{equation}
p=\det A\text{ \ \ and \ \ }q=\det A^{(1)}.
\end{equation}
From this, we observe that the order of the defect group is $\left\vert
\mathcal{C}\right\vert =\det A=p$.

Iteratively solving the group relations in equation (\ref{relator}), we also
see that either $a_{1}$ or $a_{k}$ can serve as a generator for the entire
group. That is, we have a cyclic group of order $\det A = p$. This establishes
the claim that $\mathcal{C}\simeq\mathbb{Z}_{p}$. Moreover, since our $\Gamma$
is already abelian, the abelianization is clearly $\mathbb{Z}_{p}$. Finally,
note that in the special case where the base is just a collection of $-2$
curves, we recover the special case of the defect group for the A-type $(2,0)$ theories.

\subsection{Generalized D-type Bases}

Finally, we come to the D-type bases. Recall that these are given by a
configuration of curves arranged according to the generalized Dynkin diagram:
\begin{equation}
\text{D-type Endpoint}\text{: \ \ }2,\overset{2}{n},m_{1},...,m_{l},
\end{equation}
so that the adjacency matrix takes the form:%
\begin{equation}
A=%
\begin{pmatrix}
2 &  & -1 &  & \cdots &  & \\
& 2 & -1 &  & \cdots &  & \\
-1 & -1 & n & -1 & \cdots &  & \\
&  & -1 & m_{1} & \cdots &  & \\
\vdots & \vdots & \vdots & \vdots & \ddots & \vdots & \vdots\\
&  &  &  & \cdots & m_{l-1} & -1\\
&  &  &  & \cdots & -1 & m_{l}%
\end{pmatrix}
.
\end{equation}

Contracting all of these curves leads us to the orbifold singularity
$\mathbb{C}^{2}/\Gamma$ where $\Gamma=D_{p+q,q}$ is a D-type discrete subgroup
of $U(2)$, where $p$ and $q$ are relatively prime positive integers given by
the continued fraction (see e.g. \cite{Brieskorn, Riemen:dvq, UTWOorb}):
\begin{equation}
\frac{p}{q}=(n-1)-\frac{1}{m_{1}-\frac{1}{m_{2}-...\frac{1}{m_{l}}}}.
\end{equation}
The integers $p$ and $q$ are given by the determinants of the reduced
adjacency matrices $B$, and $B^{(1)}$, the matrix obtained by deleting the
first row and column:
\begin{equation}
p=\det B\text{ \ \ and \ \ }q=\det B^{(1)},
\end{equation}
where:%
\begin{equation}
B=%
\begin{pmatrix}
n-1 & -1 & \cdots &  & \\
-1 & m_{1} & -1 &  & \\
& -1 & \ddots & -1 & \\
&  & -1 & m_{l-1} & -1\\
&  &  & -1 & m_{l}%
\end{pmatrix}
\text{, \ \ }B^{(1)}=%
\begin{pmatrix}
m_{1} & -1 &  & \\
-1 & \ddots & -1 & \\
& -1 & m_{l-1} & -1\\
&  & -1 & m_{l}%
\end{pmatrix}
.
\end{equation}

The specific orbifold group depends on whether $p$ is even or odd:%
\begin{equation}
D_{p+q,q}=\left\{
\begin{array}
[c]{l}%
\left\langle \psi_{2q},\varphi_{2p},\tau\right\rangle \text{ \ \ }p\text{
odd}\\
\left\langle \psi_{2q},\lambda_{4p}\right\rangle \text{ \ \ \ \ \ }p\text{
even}%
\end{array}
\right\}  ,
\end{equation}
where we have introduced the generators:%
\begin{align}
\psi_{k} &  =\left[
\begin{array}
[c]{cc}%
e^{2\pi i/k} & \\
& e^{-2\pi i/k}%
\end{array}
\right]  \text{, \ \ \ }\varphi_{k}=\left[
\begin{array}
[c]{cc}%
e^{2\pi i/k} & \\
& e^{2\pi i/k}%
\end{array}
\right]  \\
\tau &  =\left[
\begin{array}
[c]{cc}
& i\\
i &
\end{array}
\right]  \text{, \ \ }\lambda_{k}=\varphi_{k}\tau=\left[
\begin{array}
[c]{cc}
& ie^{2\pi i/k}\\
ie^{2\pi i/k} &
\end{array}
\right]  \text{.}%
\end{align}
As an abstract group, $D_{p+q,q}$ for $p$ odd and even is:%
\begin{equation}
D_{p+q,q}=\left\{
\begin{array}
[c]{l}%
\left\langle \psi,\varphi,\tau|\varphi^{2p}=1\text{, }\tau^{2}=\psi
^{q}=\varphi^{p}\text{, }\varphi\psi=\psi\varphi\text{, }\varphi\tau
=\tau\varphi\text{, }\psi\tau=\tau\psi^{-1}\right\rangle \text{ \ \ }p\text{
odd}\\
\left\langle \psi,\lambda|\lambda^{4p}=1\text{, }\psi^{q}=\lambda^{2p}\text{,
}\psi\lambda=\lambda\psi^{-1}\right\rangle \text{ \ \ }p\text{ even}%
\end{array}
\right\}  .
\end{equation}

Let us now turn to the abelianization of $D_{p+q,q}$:%
\begin{equation}
\text{Ab}[D_{p+q,q}]=\left\{
\begin{array}
[c]{l}%
\left\langle \psi,\varphi,\tau|\varphi^{2p}=1\text{, }\tau^{2}=\psi
^{q}=\varphi^{p}\text{, }\psi^{2}=1\right\rangle _{\text{comm}}\text{
\ \ }p\text{ odd}\\
\left\langle \psi,\lambda|\lambda^{4p}=1\text{, }\psi^{q}=\lambda^{2p}\text{,
}\psi^{2}=1\right\rangle _{\text{comm}}\text{ \ \ \ \ \ \ }p\text{ even}%
\end{array}
\right\}  .
\end{equation}
Since $p$ and $q$ are relatively prime, the pair $(p,q)$ has three
possibilities: (even,odd), (odd, odd) and (odd, even). Let us treat each of
these cases in turn.

For $(p,q)=$ $($odd,even$)$, we see that since $\psi^{2}=1$, we also have
$\psi^{q}=1$, so we also have $\tau^{2}=\psi^{q}=\varphi^{p}=1$. In other
words, the group is isomorphic to $%
\mathbb{Z}
_{2}^{(\psi)}\times%
\mathbb{Z}
_{p}^{(\varphi)}\times%
\mathbb{Z}
_{2}^{(\tau)}\simeq%
\mathbb{Z}
_{2p}\times%
\mathbb{Z}
_{2}$, where on the lefthand side we have indicated the explicit generators by a
superscript, and in the isomorphism we used the fact that $p$ is odd.

For $(p,q)=($odd,odd$)$, it is helpful to write $q=2s+1$. Then, since
$\psi^{2}=1$, we learn that $\tau^{2}=\psi=\varphi^{p}$. So in other words,
the independent elements are $\tau^{a}\varphi^{b}$ for $a=0,1$ and
$b=0,...,2p-1$, and our group is of order $4p$. Let us proceed by showing it is a cyclic group. To see this, let us show that it has an element of order $4p$. We claim it is $\tau \varphi$, indeed $(\tau \varphi)^{2p} = \tau^{2p}=\varphi^{p^2} = \varphi^p\neq1$, as $p$ is odd. Let $k$ be the smallest natural number with $(\tau \varphi)^{k}=1$: $k|4p$ but it cannot divide $2p$. As $\text{gcd}(2p,4p)=2p$, this forces $k=4p$. So in this case we find $\mathbb{Z}%
_{4p}$.

Finally, for $(p,q)=($even,odd$)$, we can again write $q=2s+1$. Then, we have
$\psi^{q}=\psi=\lambda^{2p}$, so the group is generated by $\lambda$, an
element of order $4p$.

Summarizing, we learn that the type of group is actually dictated by whether
$q$ is even or odd:%
\begin{equation}
\text{Ab}[D_{p+q,q}]=\left\{
\begin{array}
[c]{l}%
\mathbb{Z}
_{4p}\text{ \ \ \ \ \ \ \ \ \ \ }q\text{ odd}\\%
\mathbb{Z}
_{2p}\times%
\mathbb{Z}
_{2}\text{ \ \ \ }q\text{ even}%
\end{array}
\right\}  .
\end{equation}

We now show that this same structure is reproduced by a direct computation of
the defect group. To this end, we first observe that the determinant of the
adjacency matrix satisfies:%
\begin{equation}
\det A=4p.
\end{equation}
As a consequence, we always have $\left\vert \Lambda_{\text{string}}^{\ast
}/\Lambda_{\text{string}}\right\vert =4p$. To determine the defect group
$\mathcal{C}=\Lambda_{\text{string}}^{\ast}/\Lambda_{\text{string}}$, we work
in a basis where the generators are $v=(b_{1},b_{2},c,a_{1},...,a_{l})$.
Recursively solving equation (\ref{relator}), we learn that the other
generators are all obtained from appropriate powers of $a\equiv a_{l}$ and
$b\equiv b_{1}$. Moreover, since $b^{2}=(b_{2})^{2}=c$, we also find:%
\begin{equation}
a^{2p}=1\text{ \ \ and \ \ }b^{2}=a^{q}.
\end{equation}
Hence, the defect group is given by:%
\begin{equation}
\mathcal{C}=\left\langle a,b|a^{2p}=1\text{, }b^{2}=a^{q}\right\rangle
_{\text{comm}}.
\end{equation}
When $q=2s+1$, $(ba^{-s})^2 = a$, so the group is an order $4p$
cyclic group generated by the element $ba^{-s}$. On the other hand, when
$q=2s$, the group may be written as
\begin{equation}
\mathcal{C}=\left\langle a,\beta|a^{2p}=1\text{, }\beta^{2}=1\right\rangle
_{\text{comm}} \cong \mathbb{Z}_{2p} \times \mathbb{Z}_2
\end{equation}
with $\beta \equiv ba^{-s}$. Summarizing, we learn:%
\begin{equation}
\mathcal{C=}\text{ Ab}[D_{p+q,q}]=\left\{
\begin{array}
[c]{l}%
\mathbb{Z}
_{4p}\text{ \ \ \ \ \ \ \ \ \ \ }q\text{ odd}\\%
\mathbb{Z}
_{2p}\times%
\mathbb{Z}
_{2}\text{ \ \ \ }q\text{ even}%
\end{array}
\right\}  .
\end{equation}

\subsubsection*{List of Defect Groups for D-type Theories}

Let us now apply these general considerations to determine the range of
possible defect groups for the generalized D-type orbifold groups. First of
all, we recall that the generalized D-type endpoints are all of the form \cite{Heckman:2013pva}:%
\begin{equation}
2\overset{2}{3}2,\ \ 2\overset{2}{2}\underset{l\geq1}{\underbrace{2...24}%
},\ \ 2\overset{2}{2}\underset{l\geq1}{\underbrace{2...23}\text{,}%
}\ \ 2\overset{2}{2}\underset{l\geq2}{\underbrace{2...32}}.
\end{equation}
For these cases, the associated values of $p$ and $q$ are:%
\begin{align}
2\overset{2}{3}2 &  \text{:}\text{ \ \ }\frac{p}{q}=\frac{3}{2}\\
2\overset{2}{2}\underset{l\geq1}{\underbrace{2...24}} &  \text{:}\text{
\ \ }\frac{p}{q}=\frac{3}{3l+1}\\
2\overset{2}{2}\underset{l\geq1}{\underbrace{2...23}} &  \text{:}\text{
\ \ }\frac{p}{q}=\frac{2}{2l+1}\\
2\overset{2}{2}\underset{l\geq2}{\underbrace{2...32}} &  \text{:}\text{
\ \ }\frac{p}{q}=\frac{3}{3l-1}.
\end{align}
So in other words, the only defect groups for the D-type theories
are:%
\begin{equation}
C=\left\{
\begin{array}
[c]{l}%
\mathbb{Z}_{4},%
\mathbb{Z}
_{8},\text{ or }%
\mathbb{Z}
_{12}\text{ \ \ \ \ \ \ \ \ \ \ \ \ \ \ \ \ \ \ \ \ \ \ \ \ \ }l\text{ even}\\%
\mathbb{Z}
_{2}\times%
\mathbb{Z}
_{2},%
\mathbb{Z}
_{2}\times%
\mathbb{Z}
_{4},\text{ or }%
\mathbb{Z}
_{2}\times%
\mathbb{Z}
_{6}\text{ \ \ \ \ \ }l\text{ odd}%
\end{array}
\right\}  .
\end{equation}
where for completeness we have included the case of the D-type orbifold
subgroups of $SU(2)$.

\section{Compactification \label{sec:COMPACT}}

So far, our discussion has focussed on the role of the defect group in six
dimensions, and in particular how it serves to characterize discrete
three-form fluxes via the abelian group $H^{3}(M_{6},\mathcal{C})$, for an SCFT on a
six-manifold $M_{6}$. It is natural to ask how this data shows up when we
compactify a $(1,0)$ theory to lower dimensions. Much of what we say in this
section has been stated in various forms in the literature for the $(2,0)$
SCFTs. Our discussion will most closely follow that given in references \cite{Witten:2009at,Tachikawa:2013hya}.
As far as we are aware, however, the extension to $(1,0)$ theories has not been previously worked out.

Upon compactification on a $(6-d)$-dimensional Riemannian manifold $\Sigma_{d}$,
we reach a $d$-dimensional theory. The tensor multiplet will descend to a
collection of vector multiplets. For example, by compactifying on a circle,
our two-form potential converts to a standard gauge field. In the case of
compactification on a Riemann surface, each of the one-cycles similarly leads
to an additional vector multiplet. Now, in the context of 6D\ $(1,0)$ SCFTs,
there will generically be additional 6D\ vector multiplets, which upon
reduction will also contribute to the field content of the lower-dimensional
theory. Importantly, this data is independent of $\mathcal{C}$, and in
particular does not appear to lead to additional discrete flux data. We shall
therefore neglect it in what follows.

So let us now study how the defect group shows up in compactifications to lower
dimensions $d < 6$. Let us assume that some $d$-dimensional theory
is obtained by compactifying
the six-dimensional theory on a manifold $\Sigma$ of dimension $(6-d)$.
In order for the $d$-dimensional theory to have a well-defined partition function,
we must be able to assign an element $Z_{\Sigma}(M_d)$ in the partition vector
space of the six-dimensional theory for any $M_d$. Naively, when $\mathcal{C}$ is
non-trivial, this is not possible, since the vector space would typically be
multi-dimensional. However, we expect that upon specifying some additional data, a ``flux ensemble'' $F$,
it is possible to pick out (up to an overall constant of proportionality) a unique vector $Z_{\Sigma,F}(M_d)$ in the
partition vector space. Essentially, a flux ensemble $F$ provides a canonical mapping $L_F$ which
assigns to each $M_d$ a maximal isotropic subgroup $L_F(M_d)$
of $H^3 (M_d \times \Sigma,\mathcal{C})$. We can then define the
partition function $Z_{\Sigma,F}(M_d)$ to be the partition vector invariant
under the maximal isotropic subgroup $L_F (M_d)$, i.e.,
\begin{equation}
Z_{\Sigma,F}(M_d) \equiv Z^{L_F(M_d)}_{0}(M_d) \,.
\end{equation}
This picks out a unique partition function up to a constant,
since we know that for any maximal isotropic subgroup of
$H^3 (M_d \times \Sigma,\mathcal{C})$,
there is a unique ray in the partition vector space
that is invariant under it.
Since the definition is rather abstract, let us demonstrate the above by
finding the set of allowed flux ensembles $\mathcal{F} = \{ F \}$ for specific compactifications.

First of all, let us compactify on a circle to reach a 5D\ effective theory.
That is, we take the special case $M_{6}=M_{5}\times
S^{1}$. The compactification of a $(1,0)$ theory
yields an $\mathcal{N}=1$ theory in five dimensions. We study
this compactification in detail to illustrate how the partition
function is fixed using flux ensembles. Now the discrete three-form fluxes decompose as:%
\begin{equation}
H^{3}(M_{5} \times S^1,\mathcal{C})=H^{3}(M_{5},\mathcal{C})\oplus H^{2}(M_{5}%
,\mathcal{C}).
\label{decomposition}
\end{equation}
We claim that
\begin{equation}
\mathcal{F} \equiv \{ F=(G_1,G_2) :
G_1, G_2 \subset \mathcal{C},~
G_1 = G_2^\perp,~ G_2 = G_1^\perp \} \,.
\end{equation}
where
\begin{equation}
G^\perp = \{ g \in \mathcal{C} : \text{exp}({2 \pi i\langle g,g'\rangle})=1~
\text{for all $g' \in G$} \} \,.
\end{equation}
For example, when $\mathcal{C} = \mathbb{Z}_N$ generated by
the order-$N$ element $1$,
\begin{equation}
G_1 = [ p ] \cong \mathbb{Z}_q, \quad
G_2 = [ q ] \cong \mathbb{Z}_p
\end{equation}
satisfy these relations when $pq =N$. Here we have used the standard
notation where $[a]$ is the additive group generated by the element $a$.
Now given a flux ensemble $F=(G_1,G_2) \in \mathcal{F}$, the map $L_F$ is given by:
\begin{equation}
L_F (M_5) = H^3(M_5,G_1) \oplus H^2(M_5,G_2)\,.
\end{equation}
$L_F (M_5)$ is maximal and isotropic, and hence
$L_F (M_5)$ defines a unique partition vector $Z^{L_F}_0$.
We will see that the set of discrete data $\mathcal{F}$,
in the case of $(2,0)$ theories,
boils down to the choice of the gauge group of the
5D gauge theory \cite{Witten:2009at}.

To be more concrete, let us consider the case when $\mathcal{C} =\mathbb{Z}_N$.
In this case, a divisor $p$ of $N$ specifies the flux ensemble
needed to define the 5D theory:
\begin{equation}
F_{p} = ([p],[N/p]) \,.
\end{equation}
Note that $H^{3}(M_{5} \times S^1,\mathcal{C})=F_{1} \oplus F_{N}$ as
\begin{equation}
L_{F_{1}} (M_5) = H^3(M_5, \mathbb{Z}_N), \quad
L_{F_{N}} (M_5) = H^2(M_5, \mathbb{Z}_N) \,.
\end{equation}
Each $L_{F_{p}}$ with $p|N$ singles out a ray
\begin{equation}
Z_{S^1,F_{p}} = Z^{L_{F_{p}}}_0
\end{equation}
in the partition vector space of the theory. We can express the partition function
$Z_{S^1,F_{p}}$ using the basis of vectors
\begin{equation}
Z_v \equiv Z^{L_{F_{1}}}_v,\quad
v \in H^{3}(M_{5} \times S^1,\mathcal{C})/L_{F_{1}} = H^2(M_5, \mathbb{Z}_N) \,.
\label{D1}
\end{equation}
It is given by
\begin{equation}
Z_{S^1,F_{p}} =
\sum_{v \in H^2(M_5, [N/p])} Z_v \,.
\label{Zp}
\end{equation}
It is simple to check that this element of the partition vector space is
invariant under the action of the Heisenberg group operators $\Phi(v)$
for $v \in H^3(M_5,[p]) \oplus H^2(M_5,[N/p])$.

When the six-dimensional theory is an $A_{N-1}$ (2,0) theory,
$v \in H^2 (M_5,[N/p])$ are the Stiefel-Whitney classes of a $SU(N)/\mathbb{Z}_p$
gauge bundle, where $\mathbb{Z}_p = [N/p]$ is a subgroup of the
center $\mathbb{Z}_N$ of $SU(N)$. The basis vectors $Z_v$
then have the interpretation as the partition function of
a five-dimensional $SU(N)/\mathbb{Z}_p$
gauge theory restricted to gauge bundles of Stiefel-Whitney class $v$
\cite{Witten:2009at}. It follows that the partition function $Z_{S^1,F_{p}}$
is the partition function of 5D $\mathcal{N}=2$ $SU(N)/\mathbb{Z}_p$
super-Yang-Mills theory. We therefore see that the discrete data in this case
merely specifies the global structure of the gauge group of the five-dimensional
theory, as claimed. For compactifications of $(1,0)$ theories, such
an interpretation does not (yet) exist, although the partition function is
defined by the same formula (\ref{Zp}) when $\mathcal{C}=\mathbb{Z}_N$.

Next, consider compactification on $\Sigma$ a genus $g$ Riemann surface, i.e. the case
$M_{6}=M_{4}\times\Sigma$. When a $(1,0)$ theory is compactified on a torus,
it has $\mathcal{N}=2$ supersymmetry. For other compactification manifolds, the
supersymmetry of the four-dimensional effective theory is
$\mathcal{N} = 1$, i.e. four real supercharges. The discrete data that specify
the lower dimensional theory is given by a choice of a maximal isotropic subgroup $F$
of $H^{1}(\Sigma,\mathcal{C})\simeq\mathcal{C}^{2g}$ when $\Sigma$
is compact \cite{Witten:2009at, Tachikawa:2013hya}.
Given this, a maximal isotropic
subgroup $L_F (M_4)$ of $H^{3}(M_4 \times \Sigma,\mathcal{C})$ is singled out,
and hence so is a partition function, as explained in detail in \cite{Tachikawa:2013hya}.

Similar analyses should apply to compactifications of a $(1,0)$ theory on a three-manifold and
four-manifold. It would be quite interesting to work out further
physical consequences of the defect group in such cases.

\newpage

\section{Conclusions \label{sec:CONC}}

In this note we have introduced and computed the defect group of all known (and quite possibly all)
6D\ SCFTs. Quite remarkably, this data is fully captured by the abelianization
of the discrete orbifold subgroups of $U(2)$ which appear in the
classification of 6D\ SCFTs. We have determined the general pattern of
possible defect groups, and have also taken some preliminary steps in the
study of compactifying 6D SCFTs. In the remainder of this section we discuss
some avenues for further investigation.

The appearance of the abelianization of an orbifold group $\Gamma$ gives a
physical explanation for the appearance of these discrete groups in the
classification of 6D\ SCFTs. This is quite suggestive of a further role for
the theory of $\Gamma\otimes SL(2,\mathbb{Z})$ equivariant K-theory in the
study of brane charges in an F-theory compactification. Developing such a
correspondence would dovetail with the mathematical structures
observed in earlier work on the physically different case of tachyon
condensation on non-supersymmetric orbifolds (see e.g. \cite{Martinec:2002wg}%
). It would likely also point the way to a more algebraic characterization of
F-theory vacua.

We have also seen that a suitable generalization of the topological data for $(2,0)$ theories can be carried over to the case of
$(1,0)$ theories. In this spirit, it is widely suspected that there is a 7D topological field theory which governs the structure
of conformal blocks for a given $(2,0)$ theory. At a formal level, a similar structure must exist for the $(1,0)$ theories. Determining
its explicit form would be most instructive.

Finally, it is natural to ask how the topological data of the defect group carries over to those $(1,0)$ SCFTs with a holographic dual (for recent work see e.g. \cite{Gaiotto:2014lca, DelZotto:2014hpa}). In the case of the A-type $(2,0)$ theories, it is well-known how this data descends to lower-dimensional systems. The fact that there are also discrete choices in the holographic duals for compactifications of $(1,0)$ systems \cite{Apruzzi:2015wna} is quite suggestive, and would be interesting to study further.

\section*{Acknowledgements}

We thank C. C\'{o}rdova, T.T. Dumitrescu, G.W. Moore, D.R. Morrison, Y. Tachikawa, W. Taylor, A. Tsymbaliuk and C. Vafa for helpful
discussions. DP would like to thank the physics department at the University of North Carolina at
Chapel Hill and the New High Energy Theory Center at Rutgers University for their hospitality
during the completion of this work. The work of MDZ and TR is supported by NSF grant PHY-1067976. The
work of TR is also supported by the NSF GRF under DGE-1144152. The work of DP
is supported by DOE grant DE-FG02-92ER-40697.


\bibliographystyle{utphys}
\bibliography{sixDlattice}

\providecommand{\href}[2]{#2}\begingroup\raggedright\begin{thebibliography}{10}

\bibitem{VafaWitten}
C.~Vafa and E.~Witten, ``{A Strong Coupling Test of S-Duality},''
  \href{http://dx.doi.org/10.1016/0550-3213(94)90097-3}{{\em Nucl. Phys.}
  {\bfseries B431} (1994) 3--77},
\href{http://arxiv.org/abs/hep-th/9408074}{{\ttfamily arXiv:hep-th/9408074}}.

\bibitem{Ganor:1996nf}
O.~J. Ganor, ``{Six-Dimensional Tensionless Strings in the Large $N$ Limit},''
  \href{http://dx.doi.org/10.1016/S0550-3213(96)00702-X}{{\em Nucl.Phys.}
  {\bfseries B489} (1997) 95--121},
\href{http://arxiv.org/abs/hep-th/9605201}{{\ttfamily arXiv:hep-th/9605201}}.

\bibitem{Kapustin:2005py}
A.~Kapustin, ``{Wilson-'t Hooft operators in four-dimensional gauge theories
  and S-duality},'' \href{http://dx.doi.org/10.1103/PhysRevD.74.025005}{{\em
  Phys.Rev.} {\bfseries D74} (2006) 025005},
\href{http://arxiv.org/abs/hep-th/0501015}{{\ttfamily arXiv:hep-th/0501015}}.

\bibitem{Gaiotto:2010be}
D.~Gaiotto, G.~W. Moore, and A.~Neitzke, ``{Framed BPS States},''
  \href{http://dx.doi.org/10.4310/ATMP.2013.v17.n2.a1}{{\em
  Adv.Theor.Math.Phys.} {\bfseries 17} (2013) 241--397},
\href{http://arxiv.org/abs/1006.0146}{{\ttfamily arXiv:1006.0146 [hep-th]}}.

\bibitem{Aharony:2013hda}
O.~Aharony, N.~Seiberg, and Y.~Tachikawa, ``{Reading between the lines of
  four-dimensional gauge theories},''
  \href{http://dx.doi.org/10.1007/JHEP08(2013)115}{{\em JHEP} {\bfseries 1308}
  (2013) 115},
\href{http://arxiv.org/abs/1305.0318}{{\ttfamily arXiv:1305.0318}}.

\bibitem{Gaiotto:2014kfa}
D.~Gaiotto, A.~Kapustin, N.~Seiberg, and B.~Willett, ``{Generalized Global
  Symmetries},'' \href{http://dx.doi.org/10.1007/JHEP02(2015)172}{{\em JHEP}
  {\bfseries 1502} (2015) 172},
\href{http://arxiv.org/abs/1412.5148}{{\ttfamily arXiv:1412.5148 [hep-th]}}.

\bibitem{Witten:1997sc}
E.~Witten, ``{Solutions of Four-Dimensional Field Theories via $M$ Theory},''
  \href{http://dx.doi.org/10.1016/S0550-3213(97)00416-1}{{\em Nucl.Phys.}
  {\bfseries B500} (1997) 3--42},
\href{http://arxiv.org/abs/hep-th/9703166}{{\ttfamily arXiv:hep-th/9703166}}.

\bibitem{Gaiotto:2009we}
D.~Gaiotto, ``{$N=2$ dualities},''
  \href{http://dx.doi.org/10.1007/JHEP08(2012)034}{{\em JHEP} {\bfseries 1208}
  (2012) 034},
\href{http://arxiv.org/abs/0904.2715}{{\ttfamily arXiv:0904.2715 [hep-th]}}.

\bibitem{Alday:2009aq}
L.~F. Alday, D.~Gaiotto, and Y.~Tachikawa, ``{Liouville Correlation Functions
  from Four-dimensional Gauge Theories},''
  \href{http://dx.doi.org/10.1007/s11005-010-0369-5}{{\em Lett.Math.Phys.}
  {\bfseries 91} (2010) 167--197},
\href{http://arxiv.org/abs/0906.3219}{{\ttfamily arXiv:0906.3219 [hep-th]}}.

\bibitem{Gaiotto:2009hg}
D.~Gaiotto, G.~W. Moore, and A.~Neitzke, ``{Wall-crossing, Hitchin Systems, and
  the WKB Approximation},''
\href{http://arxiv.org/abs/0907.3987}{{\ttfamily arXiv:0907.3987 [hep-th]}}.

\bibitem{Witten:2009at}
E.~Witten, ``{Geometric Langlands From Six Dimensions},''
\href{http://arxiv.org/abs/0905.2720}{{\ttfamily arXiv:0905.2720 [hep-th]}}.

\bibitem{Drukker:2009tz}
N.~Drukker, D.~R. Morrison, and T.~Okuda, ``{Loop operators and S-duality from
  curves on Riemann surfaces},''
  \href{http://dx.doi.org/10.1088/1126-6708/2009/09/031}{{\em JHEP} {\bfseries
  0909} (2009) 031},
\href{http://arxiv.org/abs/0907.2593}{{\ttfamily arXiv:0907.2593 [hep-th]}}.

\bibitem{Alday:2009fs}
L.~F. Alday, D.~Gaiotto, S.~Gukov, Y.~Tachikawa, and H.~Verlinde, ``{Loop and
  surface operators in $\mathcal{N} = 2$ gauge theory and Liouville modular
  geometry},'' \href{http://dx.doi.org/10.1007/JHEP01(2010)113}{{\em JHEP}
  {\bfseries 1001} (2010) 113},
\href{http://arxiv.org/abs/0909.0945}{{\ttfamily arXiv:0909.0945 [hep-th]}}.

\bibitem{Tachikawa:2013hya}
Y.~Tachikawa, ``{On the 6d origin of discrete additional data of 4d gauge
  theories},'' \href{http://dx.doi.org/10.1007/JHEP05(2014)020}{{\em JHEP}
  {\bfseries 1405} (2014) 020},
\href{http://arxiv.org/abs/1309.0697}{{\ttfamily arXiv:1309.0697 [hep-th]}}.

\bibitem{Xie:2013vfa}
D.~Xie, ``{Aspects of line operators of class $\mathcal{S}$ theories},''
\href{http://arxiv.org/abs/1312.3371}{{\ttfamily arXiv:1312.3371 [hep-th]}}.

\bibitem{Seiberg:2011dr}
N.~Seiberg and W.~Taylor, ``{Charge Lattices and Consistency of 6D
  Supergravity},'' \href{http://dx.doi.org/10.1007/JHEP06(2011)001}{{\em JHEP}
  {\bfseries 1106} (2011) 001},
\href{http://arxiv.org/abs/1103.0019}{{\ttfamily arXiv:1103.0019 [hep-th]}}.

\bibitem{Heckman:2013pva}
J.~J. Heckman, D.~R. Morrison, and C.~Vafa, ``{On the Classification of 6D
  SCFTs and Generalized ADE Orbifolds},''
  \href{http://dx.doi.org/10.1007/JHEP05(2014)028}{{\em JHEP} {\bfseries 1405}
  (2014) 028},
\href{http://arxiv.org/abs/1312.5746}{{\ttfamily arXiv:1312.5746 [hep-th]}}.

\bibitem{Gaiotto:2014lca}
D.~Gaiotto and A.~Tomasiello, ``{Holography for (1,0) theories in six
  dimensions},'' \href{http://dx.doi.org/10.1007/JHEP12(2014)003}{{\em JHEP}
  {\bfseries 1412} (2014) 003},
\href{http://arxiv.org/abs/1404.0711}{{\ttfamily arXiv:1404.0711 [hep-th]}}.

\bibitem{DelZotto:2014hpa}
M.~Del~Zotto, J.~J. Heckman, A.~Tomasiello, and C.~Vafa, ``{6d Conformal
  Matter},'' \href{http://dx.doi.org/10.1007/JHEP02(2015)054}{{\em JHEP}
  {\bfseries 1502} (2015) 054},
\href{http://arxiv.org/abs/1407.6359}{{\ttfamily arXiv:1407.6359 [hep-th]}}.

\bibitem{Heckman:2014qba}
J.~J. Heckman, ``{More on the Matter of 6D SCFTs},''
\href{http://arxiv.org/abs/1408.0006}{{\ttfamily arXiv:1408.0006 [hep-th]}}.

\bibitem{DelZotto:2014fia}
M.~Del~Zotto, J.~J. Heckman, D.~R. Morrison, and D.~S. Park, ``{6D SCFTs and
  Gravity},''
\href{http://arxiv.org/abs/1412.6526}{{\ttfamily arXiv:1412.6526 [hep-th]}}.

\bibitem{Heckman:2015bfa}
J.~J. Heckman, D.~R. Morrison, T.~Rudelius, and C.~Vafa, ``{Atomic
  Classification of 6D SCFTs},''
\href{http://arxiv.org/abs/1502.05405}{{\ttfamily arXiv:1502.05405 [hep-th]}}.

\bibitem{Bhardwaj:2015xxa}
L.~Bhardwaj, ``{Classification of 6d $\mathcal{N} = (1,0)$ gauge theories},''
\href{http://arxiv.org/abs/1502.06594}{{\ttfamily arXiv:1502.06594 [hep-th]}}.

\bibitem{Witten:1995ex}
E.~Witten, ``{String theory dynamics in various dimensions},''
  \href{http://dx.doi.org/10.1016/0550-3213(95)00158-O}{{\em Nucl. Phys.}
  {\bfseries B443} (1995) 85--126},
\href{http://arxiv.org/abs/hep-th/9503124}{{\ttfamily arXiv:hep-th/9503124}}.

\bibitem{Witten:1995zh}
E.~Witten, ``{Some comments on string dynamics},''
\href{http://arxiv.org/abs/hep-th/9507121}{{\ttfamily arXiv:hep-th/9507121}}.

\bibitem{Strominger:1995ac}
A.~Strominger, ``{Open P-Branes},''
  \href{http://dx.doi.org/10.1016/0370-2693(96)00712-5}{{\em Phys. Lett.}
  {\bfseries B383} (1996) 44--47},
\href{http://arxiv.org/abs/hep-th/9512059}{{\ttfamily arXiv:hep-th/9512059}}.

\bibitem{WittenSmall}
E.~Witten, ``{Small Instantons in String Theory},''
  \href{http://dx.doi.org/10.1016/0550-3213(95)00625-7}{{\em Nucl. Phys.}
  {\bfseries B460} (1996) 541--559},
\href{http://arxiv.org/abs/hep-th/9511030}{{\ttfamily arXiv:hep-th/9511030}}.

\bibitem{Ganor:1996mu}
O.~J. Ganor and A.~Hanany, ``{Small $E_8$ instantons and Tensionless Non
  Critical Strings},''
  \href{http://dx.doi.org/10.1016/0550-3213(96)00243-X}{{\em Nucl. Phys.}
  {\bfseries B474} (1996) 122--140},
\href{http://arxiv.org/abs/hep-th/9602120}{{\ttfamily arXiv:hep-th/9602120}}.

\bibitem{MorrisonVafaII}
D.~R. Morrison and C.~Vafa, ``{Compactifications of F-Theory on Calabi--Yau
  Threefolds -- II},''
  \href{http://dx.doi.org/10.1016/0550-3213(96)00369-0}{{\em Nucl. Phys.}
  {\bfseries B476} (1996) 437--469},
\href{http://arxiv.org/abs/hep-th/9603161}{{\ttfamily arXiv:hep-th/9603161}}.

\bibitem{Seiberg:1996vs}
N.~Seiberg and E.~Witten, ``{Comments on String Dynamics in Six Dimensions},''
  \href{http://dx.doi.org/10.1016/0550-3213(96)00189-7}{{\em Nucl. Phys.}
  {\bfseries B471} (1996) 121--134},
\href{http://arxiv.org/abs/hep-th/9603003}{{\ttfamily arXiv:hep-th/9603003}}.

\bibitem{Seiberg:1996qx}
N.~Seiberg, ``Non-trivial fixed points of the renormalization group in six
  dimensions,'' {\em Phys. Lett. B} {\bfseries 390} (1997) 169--171,
\href{http://arxiv.org/abs/arXiv:hep-th/9609161}{{\ttfamily
  arXiv:hep-th/9609161}}.

\bibitem{Bershadsky:1996nu}
M.~Bershadsky and A.~Johansen, ``{Colliding singularities in F-theory and phase
  transitions},'' \href{http://dx.doi.org/10.1016/S0550-3213(97)00027-8}{{\em
  Nucl. Phys.} {\bfseries B489} (1997) 122--138},
\href{http://arxiv.org/abs/hep-th/9610111}{{\ttfamily arXiv:hep-th/9610111}}.

\bibitem{Brunner:1997gf}
I.~Brunner and A.~Karch, ``{Branes at orbifolds versus Hanany Witten in
  six-dimensions},''
  \href{http://dx.doi.org/10.1088/1126-6708/1998/03/003}{{\em JHEP} {\bfseries
  9803} (1998) 003},
\href{http://arxiv.org/abs/hep-th/9712143}{{\ttfamily arXiv:hep-th/9712143}}.

\bibitem{Blum:1997fw}
J.~D. Blum and K.~A. Intriligator, ``Consistency conditions for branes at
  orbifold singularities,'' {\em Nucl. Phys. B} {\bfseries 506} (1997)
  223--235,
\href{http://arxiv.org/abs/arXiv:hep-th/9705030}{{\ttfamily
  arXiv:hep-th/9705030}}.

\bibitem{Aspinwall:1997ye}
P.~S. Aspinwall and D.~R. Morrison, ``{Point-like instantons on K3
  orbifolds},'' \href{http://dx.doi.org/10.1016/S0550-3213(97)00516-6}{{\em
  Nucl.Phys.} {\bfseries B503} (1997) 533--564},
\href{http://arxiv.org/abs/hep-th/9705104}{{\ttfamily arXiv:hep-th/9705104}}.

\bibitem{Intriligator:1997dh}
K.~A. Intriligator, ``New string theories in six-dimensions via branes at
  orbifold singularities,'' {\em Adv. Theor. Math. Phys.} {\bfseries 1} (1998)
  271--282,
\href{http://arxiv.org/abs/arXiv:hep-th/9708117}{{\ttfamily
  arXiv:hep-th/9708117}}.

\bibitem{Hanany:1997gh}
A.~Hanany and A.~Zaffaroni, ``{Branes and six-dimensional supersymmetric
  theories},'' \href{http://dx.doi.org/10.1016/S0550-3213(98)00355-1}{{\em
  Nucl.Phys.} {\bfseries B529} (1998) 180--206},
\href{http://arxiv.org/abs/hep-th/9712145}{{\ttfamily arXiv:hep-th/9712145}}.

\bibitem{Witten:1996hc}
E.~Witten, ``{Five-brane effective action in M theory},''
  \href{http://dx.doi.org/10.1016/S0393-0440(97)80160-X}{{\em J.Geom.Phys.}
  {\bfseries 22} (1997) 103--133},
\href{http://arxiv.org/abs/hep-th/9610234}{{\ttfamily arXiv:hep-th/9610234}}.

\bibitem{Aharony:1998qu}
O.~Aharony and E.~Witten, ``{Anti-de Sitter space and the center of the gauge
  group},'' \href{http://dx.doi.org/10.1088/1126-6708/1998/11/018}{{\em JHEP}
  {\bfseries 9811} (1998) 018},
\href{http://arxiv.org/abs/hep-th/9807205}{{\ttfamily arXiv:hep-th/9807205}}.

\bibitem{Witten:1998wy}
E.~Witten, ``{AdS / CFT correspondence and topological field theory},''
  \href{http://dx.doi.org/10.1088/1126-6708/1998/12/012}{{\em JHEP} {\bfseries
  9812} (1998) 012},
\href{http://arxiv.org/abs/hep-th/9812012}{{\ttfamily arXiv:hep-th/9812012}}.

\bibitem{Moore:2004jv}
G.~W. Moore, ``{Anomalies, Gauss laws, and Page charges in M-theory},''
  \href{http://dx.doi.org/10.1016/j.crhy.2004.12.005}{{\em Comptes Rendus
  Physique} {\bfseries 6} (2005) 251--259},
\href{http://arxiv.org/abs/hep-th/0409158}{{\ttfamily arXiv:hep-th/0409158}}.

\bibitem{Belov:2006jd}
D.~Belov and G.~W. Moore, ``{Holographic Action for the Self-Dual Field},''
\href{http://arxiv.org/abs/hep-th/0605038}{{\ttfamily arXiv:hep-th/0605038}}.

\bibitem{Freed:2006yc}
D.~S. Freed, G.~W. Moore, and G.~Segal, ``{Heisenberg Groups and Noncommutative
  Fluxes},'' \href{http://dx.doi.org/10.1016/j.aop.2006.07.014}{{\em Annals
  Phys.} {\bfseries 322} (2007) 236--285},
\href{http://arxiv.org/abs/hep-th/0605200}{{\ttfamily arXiv:hep-th/0605200}}.

\bibitem{Witten:2007ct}
E.~Witten, ``{Conformal Field Theory In Four And Six Dimensions},''
\href{http://arxiv.org/abs/0712.0157}{{\ttfamily arXiv:0712.0157 [math.RT]}}.

\bibitem{Henningson:2010rc}
M.~Henningson, ``{The partition bundle of type $A_{N-1}$ (2, 0) theory},''
  \href{http://dx.doi.org/10.1007/JHEP04(2011)090}{{\em JHEP} {\bfseries 1104}
  (2011) 090},
\href{http://arxiv.org/abs/1012.4299}{{\ttfamily arXiv:1012.4299 [hep-th]}}.

\bibitem{Freed:2012bs}
D.~S. Freed and C.~Teleman, ``{Relative quantum field theory},''
  \href{http://dx.doi.org/10.1007/s00220-013-1880-1}{{\em Commun.Math.Phys.}
  {\bfseries 326} (2014) 459--476},
\href{http://arxiv.org/abs/1212.1692}{{\ttfamily arXiv:1212.1692 [hep-th]}}.

\bibitem{HooftI}
G.~'t~Hooft, ``{On the Phase Transition Towards Permanent Quark Confinement},''
\href{http://dx.doi.org/10.1016/0550-3213(78)90153-0}{{\em Nucl.Phys.}
  {\bfseries B138} (1978) 1}.

\bibitem{HooftII}
G.~'t~Hooft, ``{A Property of Electric and Magnetic Flux in Nonabelian Gauge
  Theories},''
\href{http://dx.doi.org/10.1016/0550-3213(79)90595-9}{{\em Nucl.Phys.}
  {\bfseries B153} (1979) 141}.

\bibitem{Henningson:2009gv}
M.~Henningson, ``{Automorphic properties of (2,0) theory on $T^6$},''
  \href{http://dx.doi.org/10.1007/JHEP01(2010)090}{{\em JHEP} {\bfseries 1001}
  (2010) 090},
\href{http://arxiv.org/abs/0911.5643}{{\ttfamily arXiv:0911.5643 [hep-th]}}.

\bibitem{BUMP}
D.~Bump, {\em Lie Groups, 2nd edition}.
\newblock Springer, 2013.

\bibitem{YELLOW}
P.~DiFrancesco, P.~Mathieu, and D.~Senechal, {\em Conformal Field Theory}.
\newblock Springer, 1997.

\bibitem{Deser:1997se}
S.~Deser, A.~Gomberoff, M.~Henneaux, and C.~Teitelboim, ``{P-brane dyons and
  electric magnetic duality},''
  \href{http://dx.doi.org/10.1016/S0550-3213(98)00179-5}{{\em Nucl.Phys.}
  {\bfseries B520} (1998) 179--204},
\href{http://arxiv.org/abs/hep-th/9712189}{{\ttfamily arXiv:hep-th/9712189}}.

\bibitem{HeisenbergGroup}
A.~Prasad and M.~K. Vemuri, ``Inductive algebras for finite {H}eisenberg
  groups,'' \href{http://dx.doi.org/10.1080/00927870902828520}{{\em Comm.
  Algebra} {\bfseries 38} no.~2, (2010) 509--514}.
  \url{http://dx.doi.org/10.1080/00927870902828520}.

\bibitem{TataTheta}
D.~Mumford, \href{http://dx.doi.org/10.1007/978-0-8176-4579-3}{{\em Tata
  lectures on theta. {III}}}, vol.~97 of {\em Progress in Mathematics}.
\newblock Birkh\"auser Boston, Inc., Boston, MA, 1991.
\newblock \url{http://dx.doi.org/10.1007/978-0-8176-4579-3}.
\newblock With the collaboration of Madhav Nori and Peter Norman.

\bibitem{Morrison:2012np}
D.~R. Morrison and W.~Taylor, ``{Classifying bases for 6D F-theory models},''
  \href{http://dx.doi.org/10.2478/s11534-012-0065-4}{{\em Centr. Eur. J. Phys.}
  {\bfseries 10} (2012) 1072--1088},
\href{http://arxiv.org/abs/1201.1943}{{\ttfamily arXiv:1201.1943 [hep-th]}}.

\bibitem{MR604577}
J.~McKay, ``Graphs, singularities, and finite groups,'' in {\em The {S}anta
  {C}ruz {C}onference on {F}inite {G}roups ({U}niv. {C}alifornia, {S}anta
  {C}ruz, {C}alif., 1979)}, vol.~37 of {\em Proc. Sympos. Pure Math.},
  pp.~183--186.
\newblock Amer. Math. Soc., Providence, R.I., 1980.

\bibitem{jung}
H.~W.~E. Jung, ``Darstellung der {F}unktionen eines algebraischen {K}{\"o}rpers
  zweier unabh{\"a}ngiger {V}er{\"a}nderlicher x, y in der {U}mgebung einer
  {S}telle x = a, y = b,'' {\em J. Reine Angew. Math.} {\bfseries 133} (1908)
  289--314.

\bibitem{MR0062842}
F.~Hirzebruch, ``\"{U}ber vierdimensionale {R}iemannsche {F}l\"achen
  mehrdeutiger analytischer {F}unktionen von zwei komplexen
  {V}er\"anderlichen,'' {\em Math. Ann.} {\bfseries 126} (1953) 1--22.

\bibitem{Brieskorn}
E.~Brieskorn, ``{Rationale singularit\"{a}ten komplexer fl\"{a}chen},'' {\em
  Invent. Math.} {\bfseries 4} (1968) 336--358.

\bibitem{Riemen:dvq}
O.~Riemenschneider, ``Deformationen von {Q}uotientensingularit\"aten (nach
  zyklischen {G}ruppen),'' {\em Math. Ann.} {\bfseries 209} (1974) 211--248.

\bibitem{UTWOorb}
O.~Iyama and M.~Wemyss, ``{The classification of special Cohen-Macaulay
  modules},'' {\em Math. Zeitsch.} {\bfseries 265} (2009) 41--83,
  \href{http://arxiv.org/abs/0809.1958}{{\ttfamily arXiv:0809.1958 [math.AG]}}.

\bibitem{Martinec:2002wg}
E.~J. Martinec and G.~W. Moore, ``{On decay of K theory},''
\href{http://arxiv.org/abs/hep-th/0212059}{{\ttfamily arXiv:hep-th/0212059}}.

\bibitem{Apruzzi:2015wna}
F.~Apruzzi, M.~Fazzi, A.~Passias, A.~Rota, and A.~Tomasiello, ``{Holographic
  compactifications of (1,0) theories from massive IIA supergravity},''
\href{http://arxiv.org/abs/1502.06616}{{\ttfamily arXiv:1502.06616 [hep-th]}}.

\end{thebibliography}\endgroup

\end{document}